\documentclass[twocolumn]{article}
\usepackage{graphicx}

\begin{document}

\title{On the change of latitude of Arctic East Siberia at the end of the Pleistocene} 

\author{W. Woelfli\footnote{ Institute for Particle Physics, ETHZ H\"onggerberg, CH-8093 Z\"urich, Switzerland (Prof.~emerit.); e-mail: woelfli@phys.ethz.ch .} \ and
W. Baltensperger\footnote{Centro Brasileiro de Pesquisas F\'\i sicas, Rua Dr.\thinspace Xavier Sigaud,150, 222\thinspace 90 Rio de Janeiro, Brazil;\qquad \qquad e-mail: baltens@cbpf.br}}
\date{12 April 2007}
\maketitle

\noindent\emph{Mammoths lived in Arctic East Siberia. In this region there is not sufficient sunlight over the year for the growth of the plants on which these animals feed. Therefore the latitude of this region was lower before the end of the Pleistocene. As the cause of this geographic pole shift, we postulate a massive object, which moved in an extremely eccentric orbit and was hot from tidal work and solar radiation. Evaporation produced a disk-shaped cloud of ions around the Sun. This cloud partially shielded the solar radiation, producing the cold and warm periods that characterize the Pleistocene. The shielding depends on the inclination of Earth's orbit, which has a period of 100'000 years. The cloud builds up to a density at which inelastic particle collisions induce its collapse The resulting near-periodic time dependence resembles that of  Dansgaard-Oeschger events. During cold periods fine grained inclusions were deposited into the ice. The Pleistocene ended when the massive object had a close encounter with the Earth, which suffered a one per mil stretching deformation. While the deformation relaxed to an equilibrium shape in one to several years, the globe turned relative to the rotation axis: The North Pole moved from Greenland to the Arctic Sea. The massive object was torn to pieces, which evaporated.}

\section{Revealing evidence}
Remains of mammoths have been found in East Siberia in regions with a high latitude, where at present these herbivores could not exist. In East Siberia herds of mammoths grazed within the arctic circle, even on islands in the Arctic Sea, which were connected with the mainland during the glacial periods. These facts were first reported in the Soviet literature and are confirmed by current investigations \cite{Schirrmeister, Orlova, Bocherens}. The data are complex, since the temperature varied between stadials and interstadials, but consistently this area was not ice covered. 

The yearly insolation decreases with increasing latitude. The present distribution of the flora on the globe suggests that in arctic regions  the yearly insolation is insufficient for steppe plants. A limiting situation exists in the Wrangle Islands, where in some favourable habitats certain depauperated relicts of the Pleistocene grassland supported the persistence of dwarf mammoths into the Holocene \cite{Vartanian}. Floral reconstructions for the Late Pleistocene have also been made on the basis of fossil beetles in arctic East Beringa \cite{Alfimov}. The present paper is written with the assumption that beyond a limit given essentially by the arctic circle the insufficient yearly insolation inhibits the growth of the steppe plants indispensable for mammoths. Why then could mammoths exist in arctic areas? Since this depends on the yearly insolation, there is just one answer: these regions had a lower latitude in the Pleistocene. 

Let us suppose that the North pole was at the center of the known ice cover of the Last Glaciation. This is situated in Greenland, about $18^\circ$ apart from the present North pole. Its longitude is less certain, since it depends on the thermal influence attributed to the Atlantic ocean. A model study of the Pleistocene climate with geographically shifted poles could be revealing. The geographic consequences of a polar shift are best visualized using a globe.  The latitude is always (90$^\circ$ - angular distance from the pole).  Some places on the great circle through the old and the new positions of the poles suffered the full 18$^\circ$ shift. The Lena River in Siberia moved  18$^\circ$ north, while the latitudes in Australia decreased approximately by this amount. Bolivia moved away from the equator (tropical $\rightarrow$ arid), while the Northern Amazon region shifted to the equator (arid $\rightarrow$ tropical). The latitudes on the US-East coast and in West Europe were higher in the Pleistocene, and those of Alaska slightly lower. Since the North pole moved from Greenland into the Arctic Sea while the South Pole was displaced within Antarctica, the climate changes were larger on the Northern hemisphere.

The evidence of mammoths in Arctic East Siberia is not just one more in a multitude of unexplained facts, since it contains an aspect that we understand. These regions necessarily received more sunlight in the Pleistocene than at present.  Thus, the latitude of arctic East Siberia was lower than it is now. The globe has been turned with respect to its rotation axis. If the idea of the geographic shift of the poles appears out of the ordinary, then so are the empirical facts. (They are of a kind similar to the data on photoemission at the beginning of the last century.  The Maxwell equations then continued to be valid, however,  something additional [i.e.~quantization] was necessary for the explanation.) In the present case the Milankovitch theory continues to be valid, but without a lower latitude of East Siberia during the Pleistocene the facts cannot be understood. The evidence imposes a conclusion which adds a basic assumption i.e.~a shift of latitude of Arctic East Siberia. Since this geographic pole shift took place, there must be at least one possible scenario to produce it. In the following paragraphs we attempt to find such a course of events. This involves complex processes for which we can only provide simple estimates. If more elaborate future studies will show that our estimates are wrong,  this means that the real scenario was different, but not that the shift of the poles did not occur.

The problem discussed here has a long history. At the end of the 19$^{th}$ century, several geographers postulated a polar shift on the basis of the asymmetry of the observed Pleistocene maximum glaciation. Careful studies by G.H. Darwin, J.C. Maxwell,  G.V. Schiaparelli and by W. Thomson led to the conclusion that the required rapid polar shift was impossible. The verdict of these eminent scientists appeared definite. At that time, condensed matter was considered to be either solid or liquid. During the last century the concept of plastic behaviour was conceived, which opened a new range of possible relaxation times for deformations. It may have remained unnoticed that this is of decisive importance for the problem of a rapid geographic polar shift. Without the polar shift, the problem of mammoths in polar regions would remain unsolved.

\section{Geographic polar shift}

A rapid geographical shift of the poles is physically possible \cite{Gold}. At present, the Earth is in hydrostatic equilibrium. Since it rotates, its radius is larger at the equator (by 21 km) than at the poles. The rotational motion of an object is governed by its inertial tensor. In a coordinate system fixed to the object and with the origin at the center of mass, this tensor is obtained by an integration over the density times a bilinear expression of the cartesian components. At present, due to the equatorial bulge, one of the main axes of Earth's inertial tensor is longer than the other two, and its direction coincides with that of the rotation axis. This is a stable situation. For a polar shift, a further deformation of the Earth is required. The ensuing motion leads to new geographic positions of the North and South Poles. During the shift, the direction of the angular momentum vector remains strictly fixed relative to the stars, as required by conservation laws. What turns is the globe relative to the rotation axis. 

Suppose the Earth gets deformed: some of its mass is displaced to an oblique direction. This produces an inertial tensor with a new main axis, which deviates  from the rotation axis. Then, as seen from the globe (i.e.~geographically), the rotation axis will move around ("precess" around) this main axis. Actually, on a minute scale such a precession is observed on the present Earth (Chandler precession). A full turn of the precession takes about 400 days. This period is determined by the equatorial bulge. Its order of magnitude will be of prime importance in the discussion of the polar shift. The shape of a deformed Earth relaxes to a new hydrostatic equilibrium. This brings the precession to an end. In the final situation there is again an equatorial bulge around the new geographic position of the rotation axis. The poles have shifted geographically, however, what turned in space is the globe. 

If global deformations relax in a time short compared to about 200 days (half a precession cycle), the movement stops quickly and the pole shift remains insignificant. This would happen in the case of an elastic deformation of a solid Earth, since the changes of the deformation occur with the speed of sound. Similarly, for a model of a liquid Earth, the backflow of matter (over distances of 10 km) is expected to occur within days at most. Historically, at the end of the 19$^{th}$ century, these were the only known states of condensed matter. Thus, a polar shift seemed to be impossible \cite{Hapgood}. In the last century, plastic materials with wide ranges of relaxation times were investigated. The idea that a global deformation of the Earth relaxes in several years appears plausible. A simple calculation  of a geographic polar shift with just one assumed relaxation time of 1000 days is given in the Appendix of 
Ref.~\cite{Woelfli2002}. The result is a decreasing precession of the rotation axis around the main axis, which itself moves on a small spiral. Of course, a study of the motion with a detailed model of the Earth would be very significant.

\vspace{0.5cm}\section{Cause of the deformation}

The polar shift requires a displacement of mass on Earth at the end of the Pleistocene. What mechanism could produce this? Hapgood \cite{Hapgood} proposed that the ice on Antarctica could become unstable and drift away from the South Pole due to the centrifugal force. For an appreciable shift, the ice would have to move several tens of degrees latitude and add to the mass of a continent rather than float. Even then, this displacement of mass would actually not suffice for the required shift. A related idea might consider a displacement of Earth's nucleus from the centre by centrifugal forces. Again, this displacement would have to be large, and it is incompatible with the present centered position of the nucleus.  

A very efficient deformation is a stretching of the globe in a direction oblique to the poles. A volume flow over distances of the order of the stretching amplitude can create a mass difference at the surface far from the rotation axis. For the pole shift considered, the required stretching amplitude is  $ 6.5$ km on each side \cite{Woelfli2002}. How could this one per mil stretching in a direction $30^\circ$ from the poles occur? If a massive object passed near the Earth, it would create a tidal force. Since (for large distances) tidal forces vary with the third reciprocal power of the distance to Earth, the  Moon brought 20 times closer would produce a large but still insufficient tidal effect. A close passage of a mass about ten times larger is required (approximately the mass of Mars) \cite{Woelfli1999}. It is reasonable to assume a planetary speed for this object, say a relative velocity to Earth of about 40 km/s. The close distance then lasts about 10 minutes only. The process of deformation is therefore highly dynamic. Only an elaborate study of this process could give reliable  numbers. The required one per mil value of the stretching is the deformation with a prolonged relaxation of at least a hundred days. Any deformation that decays more rapidly would be additional. The global deformation is catastrophic, although compatible with the continuation of life on Earth. Nevertheless, many large vertebrate species are known to have become extinguished at the end of the Pleistocene \cite{Martin,Orlova,Bocherens}.  

\section{Bound massive object Z}

What was this massive object, that passed near the Earth? Certainly not one of the present planets, since these have orbits that do not pass through Earth's distance from the Sun. The object involved in the near collision must afterwards have been in an orbit, which crosses that of Earth, and the Holocene was much too short for a major readjustment due to couplings with Jupiter and other planets. One might think of a massive object which happened to travel through the planetary system. This might occur as a rare event. However, it is improbable that at this occasion the object comes close to Earth. For this reason the object has to be in a bound orbit that  crosses Earth's distance from the Sun. The chance that a passage through the surface of the sphere at Earth's distance, $R_E=150$ Mio km, happens within a range of $r=20\thinspace 000$ km  is only $\pi r^2/(4\pi R_E^2)=4\cdot 10^{-9}$. Thus it is even unlikely that a narrow encounter occurs within the time in which an object under the influence of Jupiter remains  in an exotic orbit, typically a few million years. Therefore we assume that the orbital plane of this object, henceforth called Z, is restricted to a small angle (say $1^\circ$) with the invariant plane (perpendicular to the total angular momentum of the planetary system). For example Z might have been a moon of Jupiter which got loose. 

The larger the distance of an approach between Z and Earth, the  more frequent it is.  Passages of Z nearer than about the distance Earth-Moon create dramatic  earthquakes. They may well  be the trigger of Heinrich events \cite{Heinrich}. They might also perturb Moon's orbit. For a rough numerical estimate let us assume a relative velocity between Z and Earth of 40 km/s. A passage of Z through Moon's orbit then lasts 20\thinspace 000 s, a time during which Moon's velocity normally changes its direction by $3^\circ$. During the passage Earth's mass is effectively increased by that of Z, i.e. by 10 \%. Then, since the orbital angular velocity is proportional to the square root of the central mass, the Moon suffers an additional change of direction of $0.15^\circ $ only. Thus except for rare cases of a narrow approach between Z and Moon, the perturbation of Moon's orbit is compatible with its present eccentricity.  Nevertheless, Moon's rotation may not have remained synchronous with its orbital motion. However, the rotation relaxes by tidal friction in a time shorter than the Holocene \cite{Nufer}.

The orbital parameters of Z are not known, but restricted by three conditions. The aphelion lies beyond Earth's orbit. Its value together with the time dependent inclination of Z to the ecliptic should allow a close approach to Earth within some million years. We shall see that the perihelion distance must be small enough, so that the object is hot.
In numerical estimates we often used $4\cdot 10^9$ m for the perihelion distance and $1.5\cdot 10^{11}$ m for the semi-axis of the ellipse. This corresponds to an eccentricity $\epsilon = 0.973$.

\section{Disappearance of Z}

Evidently, at present Z does not exist. How could it disappear within the Holocene?  Only the Sun could accomplish this. Z had to be in a special situation before the pole shift, i.e. during the Pleistocene. Necessarily, Z had to move in an extremely eccentric orbit, with a perihelion distance barely compatible with its existence. Each time Z passed through the perihelion, it was heated inside by tidal deformation and on the surface by solar radiation. Z was liquid and had a shining surface.

Since Z must disappear during the Holocene, it is almost indispensable that Z is torn to pieces during the narrow passage. For this Z must be much smaller than Earth, so that the tidal forces produced by Earth on Z are larger than those by Z on Earth. The  pressure  release in the hot interior may have  further promoted the breakup. The condition that Z had at least 1/10 of the mass of Earth, so that it could deform the Earth as required by the polar shift, together with the condition that it was much lighter than Earth, imposed by the breakup, determine the size of Z surprisingly well. Z was about Mars-sized. Again, these considerations deserve detailed studies.
  
For an evaporation from Z, the particles have to surmount their escape energy. The escape speed of Mars is 5.02 km/s. As an example, the kinetic energy of an Oxygen atom (as the most frequent atom on a dense planet) with that speed is 2.1 eV. If half the molecular binding energy of an O$_2$ molecule is included in the cost to produce an evaporated O-atom, it turns out that it takes less energy (4.2 eV for O$_2$ versus 4.7 eV for O) to evaporate the molecule than the single atom. However, note that this holds, since O$_2$ is a fairly light molecule. In most cases, atomic evaporation prevails. In a theory of evaporation, the Boltzmann factor $\exp  [-E/(k_BT)]$ plays a dominant role, where $E$ is the energy necessary to liberate a particle, $T$ the temperature and $k_B$ the Boltzmann constant. Let us assume $T=1500$ K on the surface of Z near the perihelion. For $E$ we use the escape energy for an Oxygen molecule from Z, i.e. $E_1 = 4.2$ eV. If Z breaks into $n$ equal parts Z$_n$, each has $1/n$ the mass of Z, while its radius is reduced by $(1/n)^{1/3}$ at most.  Therefore the  escape energy  from a fragment satisfies $E_n  \leq E_1/n^{2/3}$, and the ratio between the Boltzmann factors for Z$_n$ and  Z becomes 
\begin{equation}
e^{{E_1-E_n\over k_BT}} \geq \left\{ \begin{array}{ll}2\cdot 10^5 &\mbox{ for } n=2\\2\cdot 10^7 &\mbox{ for } n=3\end{array}\right. 
\end{equation}
Thus the splitting of Z into two or more parts results in an enormous increase of the Boltzmann factor. Mostly for this reason, we expect a dramatic increase of the evaporation rate after the polar shift. From the fractions Z$_n$, molecules and clusters evaporate. Furthermore, since the near-collision  between Z and Earth dissipates energy, it is likely that the  perihelion distances of the parts Z$_n$ are reduced. If their masses diminish sizeably in the following 1000 years, then the complete evaporation within the Holocene results. Obviously, these considerations are preliminary. They indicate the possibility that Z can disappear within the Holocene, provided that it was already evaporating during the Pleistocene. 

In principle, Z could vanish in a different way. Since it was in an extremely eccentric orbit, there is a certain probability that during the narrow encounter it lost its small angular momentum and afterwards dropped into the Sun. However, as Bill Napier pointed out in a private communication,  as a result of the attraction to the Sun, a Mars-sized object would introduce a kinetic energy equivalent to the  solar radiation of three years, and furthermore, the shock might perturb the delicate equilibrium in the Sun's innermost parts and activate an increase of the nuclear reaction. For these reasons, we favour processes which gradually add the material of Z to the Sun  over many years.

A further possibility is that Z is expelled from the planetary system by the time dependent gravitational field if Jupiter. However, in the short time of the Holocene this could only occur, if the new orbit of Z happened to be in a precise resonance with that of Jupiter or if Z had a close encounter with Jupiter. This cannot be excluded, but it will have a smaller probability than the evaporation of the fractions of Z.

This assumed scenario for the Pleistocene could not have occurred several times during the existence of the planetary system without a collision of Z with one of the inner planets. The Pleistocene ice age era was a rare, if not unique, period in Earth's history.  Other types of ice ages may have existed  as a result of the slow movements of the continents. If these were joined to one block, Earth's rotation is stable when this supercontinent is centered around a pole. Plausibly, in this situation the whole continent is covered by ice. 

\section{Traces of the cloud in polar ice cores} 

The evaporation from Z creates a cloud.  At times this may screen the Earth from solar radiation and thereby produce cold periods. Muller and MacDonald \cite{Muller1995,Muller1997} have postulated a shielding of the Sun by an interplanetary cloud as the cause of the cold periods, since this depends on Earth's inclination, which has a 100 kyr period. Changes of the inclination could be much more effective than variations of the eccentricity, which have about the same period. 

Some of the material of the cloud may reach the Earth. Inclusions of non-solvable matter in ice-cores from Greenland and Antarctica have been studied extensively. It is a remarkable fact that the impurity concentrations in the ice cores of Greenland \cite{Mayewski} and of Antarctica \cite{EPICA2004}  are sharply peaked during cold periods. Densities in cold periods surpass values for interstadials by two orders of magnitude. It has been argued that in cold periods the transport of dust in Earth's atmosphere may be higher than in warm periods. However, the size of the effect suggests a more intrinsic connection between cold periods and inclusions. 

The analysis of the grain sizes of the inclusions by J.P. Steffensen \cite{Steffensen} (see Fig.~2) and B. Delmonte et al.~\cite{DelmontePetit} showed that these are composed of a large-particle fraction and a distribution of fine grains with diameters between 1 and 4 $\mu$m. The origin of the large grains has been clearly determined. The large particles from Groenland come from the Gobi desert \cite{Biscaye} and those of Antarctica from Patagonia \cite{Delmonte}. The size distribution of the large-particle fraction differs from one stadial to another. This is likely to depend on details of the storms that transported the grains. Also the contribution of the large-particle fraction to the total mass of the inclusions depends appreciably on the cold event (see Table 1 of Ref. \cite{Steffensen}. It is at most 16.5 \% ('Pre-Eemian "warm"') and can be as small as 1 \% ('Post-Eemian "cold" (2)'). In all stadials the mass of the fine grains dominate. The size distribution of the fine grains has a nearly identical shape in different cold periods with a maximum at a radius near 1 $\mu$m. This indicates that the fine grain distributions do not depend on the wind.  
The ultraviolet part of the solar radiation may ionize all atoms of the cloud. In this case single ions enter the upper atmosphere of the Earth. They will form molecules and assemble to clusters before these reach the ground. Such processes are not primarily dependent on the wind.

\section{Types of clouds}
Z evaporates its material close to the perihelion. There the velocity of Z is much larger than thermal particle velocities. Hence the initial conditions of the particle motions correspond to those of Z. However, each particle is subject to its specific light force due to the solar radiation. If the first excitation energy of the particle is larger than about 10 ev, the repulsive light force is expected to be weaker than the gravitational attraction to the Sun. This includes almost all ions, some atoms, but no molecules. In these cases bound orbits exist. In general, an evaporated particle may have a hyperbolic or elliptic orbit. 

The properties of the cloud determine the most important consequences of this model, since the cloud can partially shield  the solar radiation from  Earth, which becomes the prime reason for the glaciations  during the Pleistocene. The cloud is very complex. It involves particles,  each with its light force, plasma properties and possibly magnetic and electric fields \cite{Meyer-Vernet}. The cloud receives particles that evaporated from Z, which itself is in a time-dependent orbit.  The orbital and spin periods of Z appear in the evaporation. Furthermore, the dynamics of the cloud itself produces  time dependencies. In view of these problems statements regarding the spatial extent of the cloud and the amount of material lost to outer space are difficult. 

A first version of this paper \cite{Woelfli2006} was written under the impression, that a terrestrial origin of the inclusions in bore ice had been demonstrated. Therefore the question arouse, whether a cloud that does not reach Earth´s distance from the Sun was possible.  This type of cloud must be sufficiently dense, so that an evaporated particle makes inelastic collisions with other particles. The emitted photons reduce the energy of the particles, but practically not their angular momentum. The resulting cloud is circular and extends to about twice the perihelion distance of Z. The particles spiral into the Sun under the influence of the Poynting-Robertson drag \cite{Gustafson}. Once this time independent, small cloud exists, it may sustain itself. However, it is unclear how it could be created initially. 

It cannot be stressed enough that the dynamics of the cloud  may turn out to be a very complex theoretical problem. It is the basis for understanding the temperature changes of the ice age era. 

\section{Time dependent cloud}

Without a cloud, the evaporated particles are not stopped by collisions. If they are ions, any electric or magnetic field could modify their path. The orbits of the individual particles will be spread over rather vast space so that initially particles do not collide.  Their lifetime is limited by Poynting-Robertson drag (see e.g. Eq.~[11] of Ref.~\cite{Woelfli2006} valid for circular  motion). For an estimate,  using the mass of an O-atom, and an orbital radius of $1.5\cdot 10^{11}$ m (Earth's distance from the Sun), the lifetime becomes ${400\over f}$  years,  where $f$ is the absolute value of the ratio between the light force and the gravitational attraction to the Sun. For ions we expect $f\ll 1$. This lifetime increases with the square of the size of the orbit.  

As the influx of particles continues, the density of the cloud increases. At a certain point, energy loss by inelastic collisions becomes appreciable. This reduces the relative velocity between two colliding particles so that their orbits become more similar. The cloud becomes more disk like and its radius diminishes. As the volume of the cloud shrinks, collisions become more frequent. The resulting transition from individual motion to collision-dominated collapse of the cloud can only occur in a time shorter than the Poynting-Robertson lifetime. As the particles  move closer to the Sun, the amount of light scattered per particle increases. In the endphase of the cloud, the particles have aligned motions in small orbits, and they spiral into the Sun due to Poynting-Robertson drag. Thus the initial dilute cloud gradually becomes denser until collisions induce a collapse and lead to the elimination of the particles by Poynting-Robertson drag. If Earth's orbit lies near the midplane of the cloud, which is presumably close to the invariant plane of the planetary system, the insolation is diminished \cite{Muller1995, Muller1997}. With this cloud the slowly growing shielding is followed by a rather sudden return to the full solar insolation. This will be followed by another buildup of a cloud. Such a time dependence of the shielding reproduces characteristic features of the Dansgaard-Oeschger temperature peaks. 21 such events were counted  between 90 and 11.5 kyr BP \cite{GRIP,North}, which corresponds to 4000 yr as the average spacing between two events. More detailed data reveal a period of 1470 yr \cite{Rahmstorf}. Both periods are too short to be connected to Milankovitch modifications of orbits. On the other hand the particles of the time dependent cloud are most probably ions. Their values $f$ are sufficiently small, so that the Poynting-Robertson lifetime is longer than the Dansgaard-Oeschger periods. Therefore they can be determined by the intrinsic time dependence of the cloud. An important information comes from the new ice-bore EDML in Antarctica, which covers a range of 150 kyr. All the Daansgard-Oeschger events recorded in Greenland are also present in Antarctica. (Fig. 1 a and b of Ref. \cite{Epica2006}). Tentatively, we explain the Dansgaard-Oeschger temperature variations with  sequences of formations and collapses of a screening cloud. 

Extensive studies will be necessary to establish the conditions under which either the small cloud or the time dependent cloud is real. We are aware that we left aside plasma properties, electric and magnetic fields and the question, whether energy loss via gas discharges might play a role. From the time dependent cloud a particle flux into Earth's atmosphere seems unavoidable. It would be important to establish upper bounds for extraterrestrial components in ice cores. In our model the cloud was produced by an evaporation that was limited by the gravitational attraction to the hot object Z. This acts like a mass dependent fractional distillation. Thus the material of the cloud must show isotope effects: the light isotopes must be enhanced. The validity of this prediction could be settled by measuring the distribution of the three naturally occurring isotopes of Magnesium taken from the small grained inclusion of a cold period.

\section{Unavoidable ice age era}

In this model, Z is necessary for the polar shift, but Z does not exist any more. This is indeed a strong condition for the scenario. Clearly, Z had to be hot and emitting material already before the polar shift. This period lasted a few million cycles, the estimated number of probable trials for a narrow approach  to Earth. During this time Earth's climate was influenced by the screening of the solar radiation by the  gas cloud. The existence of a polar shift at the end of the Pleistocene, connected with the condition, that the culprit has disappeared, leads to the prediction of a variable cold period lasting a few million years. The  Milankovitch theory without polar shift predicts variations in insolation during an unlimited time forward and backward. It cannot cope with the observed fact that before about 3 million years BP the climate was warm with only small fluctuations \cite{Tiedemann} similar to that of the Holocene. 

\section{Cloud after the pole shift}

The evaporation rate increased enormously after the narrow encounter. What happened with the cloud? The pole shift occurred towards the end of the Pleistocene, probably before the Younger Dryas, which was a cold period, but not distinctly colder than previous periods. What happened? The high velocities of Earth and Z were determined by their motion in the Sun's large gravitational field.  If their relative velocity was of the order of 40 km/s, then the momentum transfer during the narrow encounter produced a modest scattering of Z  by an angle of a few degrees. Before, the orbital planes of Earth and Z probably almost coincided. After the scattering, the orbital plane of Z likely was rotated by a few degrees around an axis that passed through the location of the close encounter and the Sun. This tilt saved the Earth from an extreme shielding of the solar radiation. The evaporated material (mostly initially in molecular form) moved outside the existing disk of ionic particles. Either the new material was blown out of the planetary system by radiation pressure, or, if a new cloud of ions established itself in the tilted plane, this shielding affected the Earth only twice a year, i.e. when the Earth crossed the plane of the cloud. Again, more detailed modeling might establish probabilities of various occurrences. It seems likely that the scattering of the fractions of Z during the narrow encounter saved the Earth from dramatic shielding of solar radiation and influx of material into Earth's atmosphere.

\section{Conclusions}

Remains of mammoths in arctic Siberia, where there is not enough sunlight per year to grow the plants that feed these animals, indicate, that the latitude of this region  was lower in the Pleistocene. Therefore a scenario should exist, which leads to this geographic polar shift. The paper describes a proposal which involves a massive object Z. In a close encounter between Z and the Earth the tidal force created a 1 per mil extensional deformation on the Earth. While this deformation relaxed to a new equilibrium shape in a time of at least 200 days, the precessional motion of the globe resulted in the geographic polar shift. Since Z  does not exist any more, it must have moved in an extremely eccentric orbit, so that it was hot and evaporating. It produced a disk-shaped gas cloud around the Sun which partially shielded the solar radiation. The resulting glaciation on Earth depended on the inclination of Earth's orbit relative to the invariant plane, which has a period of about 100 kyr. The probable cloud structure is time dependent: its density increases until inelastic collisions induce its collapse. The resulting time dependence of the screening explains the Dansgaard-Oeschger temperature peaks. The Heinrich events might result from earthquakes created by passages of Z near Earth up to about Moon's distance.  During the  polar shift event the Earth must have torn Z  to pieces which  evaporated subsequently. Such a  close encounter is expected to occur only once in several millions of orbital motions of Z. Therefore the model correctly predicts a duration  for the era of glaciations of the order of the length of the Pleistocene. 

We conclude that it is possible to construct a scenario which leads to the required geographic polar shift. The resulting model has few free parameters and yet it describes the dominant observed features of the Pleistocene. It involves complex problems  which we have treated only rudimentarily. More elaborate studies would be very valuable. In the case that these contradict the given estimates, they could primarily question this particular scenario, rather than the evidence that a pole shift has occurred.

\end{document}